\documentstyle[12pt,plenum]{book}

\def\real{{\rm I\kern-.2em R}}
\def\complex{\kern.1em{\raise.47ex\hbox{
         $\scriptscriptstyle |$}}\kern-.40em{\rm C}}
\def\integer{{\rm Z\kern-.32em Z}}

\begin{document}
\chapter{TOPOLOGICAL FEATURES OF THE MAGNETIC RESPONSE IN INHOMOGENEOUS 
MAGNETIC FIELDS}

\author{E.~Akkermans\refnote{1} and R.~Narevich\refnote{1,}\refnote{2}}

\affiliation{\affnote{1}Department of Physics, Technion\\ 
32000 Haifa, Israel\\
\affnote{2}Department of Physics, University of Maryland\\
College Park, MD 20742, USA}


\begin{verse}
\small 
\noindent
{\bf Abstract} 

\noindent
We present topological features of 
the magnetic response (orbital and spin) of a two-dimensional
non interacting electron gas due to inhomogeneous applied magnetic fields.
These 
issues are analysed from the point of view of the Index theory with a special
emphasis on
the non perturbative aspects of this response. The limiting case of a 
Aharonov-Bohm magnetic flux line is studied in details and the 
results are extended to more 
general situations.

\end{verse}



\section{INTRODUCTION}


The aim of this paper is to discuss some features of the magnetic
 response 
of a degenerate two-dimensional electron gas in the non interacting limit.
 Although 
there is a vast literature devoted to that subject, 
we would like to present a different point of view which may help to bring new
 results in some of the left open issues. 
In those systems, it is quite common to discuss separately the two  
components (orbital and spin) of the magnetic response.
Very often, there is indeed a clear 
dichotomy between orbital and spin effects which might be due to some
 physical constraint
(e.g. full spin polarization in a strong magnetic field) or to the
 independence of the two components of the response, for instance for a
 homogeneous 
magnetic field where (neglecting the spin-orbit coupling)
 both the orbital and Zeeman parts in the Hamiltonian 
 do commute. Recently, it was noticed by a number of authors that more
 complicated 
situations for instance
 Dirac fermions in a random field\refnote{\cite{goldbart,wen}}
may lead to new and unexpected effects:
transition between localized and extended states, multifractal
 structure, etc. We do 
not want to discuss here the richness of some specific model but
 instead to present 
some features of the total magnetic response in 
inhomogeneous 
magnetic fields shared by most of these models in the general case
 where orbital and 
spin effects cannot be simply disentangled. These features as we shall see are 
 essentially non perturbative and do require for their study some new tools 
imported from the Index theory of elliptic operators\refnote{\cite{book}}.

The outline of this paper is as follows. In the remaining part of the 
Introduction, we shall set up a general form for the Hamiltonian we aim
 to study. 
Then, we shall discuss its factorizability and define the associated Index.
In part 2, we present a detailed study of the magnetic response to a 
Aharonov-Bohm 
flux line as a limiting case of inhomogeneous field.
 This will be the opportunity to discuss the physical 
meaning of the Index and its relation to the magnetic response. 
In part 3, some properties of the associated Heat Kernel
 are outlined. Then in part 4, those results are extended
 to other systems, and a relation between the spin magnetization
and the Index is given.

We start writing a general expression for the magnetic Schr\"{o}dinger 
Hamiltonian for a single electron in the two-dimensional plane submitted to a 
perpendicular magnetic field of strength $B(r)$. We choose the
 gauge ${\rm div} {\vec A}
=0$ such that the vector potential ${\vec A}(x,y)$ obeys the two equations:
\begin{eqnarray}
 \left\{\begin{array}{ll}
{\partial_x}{A_x} + {\partial_y}{A_y}=0 \\
{\partial_x}{A_y} - {\partial_y}{A_x}= B(r)
\end{array}
\right. 
\end{eqnarray}
A solution of (1) is 
$({A_x},{A_y}) = {{\Phi(r)} \over {2 \pi {r^2}}} (-y,x)$,
and the function $\Phi (r)$ is related to
 the magnetic field by 
$B(r)= {1 \over {2\pi r }}{\partial_r}\Phi(r)$ such that the Schr\"{o}dinger
 Hamiltonian is 
\begin{equation}
{{2m} \over {\hbar^2}} H= -{1 \over r} {\partial_r}(r{\partial_r})
+({i \over r}{\partial_\theta} + {\phi(r) \over r}{)^2}
\end{equation}
where $\phi(r) \equiv {e \over hc}\Phi(r) = {\Phi(r) \over {\Phi_0}}$. This 
Hamiltonian can be expressed as well in terms of the two (formally)
 self-adjoint 
first order differential operators $D$ and $D^\dagger$ as 
\begin{equation}
{{2m} \over {\hbar^2}} H= D{D^\dagger} - {2\pi \over {\Phi_0}}B(r)
\end{equation}
where $D= {e^{i\theta}}({\partial_r} + \hat{J})$ and ${D^\dagger}= 
{e^{-i\theta}}(-{\partial_r} + \hat{J})$ such that
 $[D,{D^\dagger}]= 4\pi B(r)$. The operator 
 $\hat{J}= 
{i \over r}{\partial_\theta} + {\phi(r) \over r}$ describes
 the azimuthal current. 

The Zeeman term describing the coupling of the electron spin to the magnetic
 field is given by ${H_s}= -{1 \over 2} g
 {\mu_B}{\sigma_z}B(r)$,
 where ${\mu_B} = {e\hbar \over 2mc}$ is the Bohr magneton,
 $g$ is the
 gyromagnetic
ratio we shall take equal to 2 and ${\sigma_z}$ is the Pauli matrix. The total
 Pauli Hamiltonian (in appropriate units) is
\begin{equation}
{{2m} \over {\hbar^2}}{H_P}= D{D^\dagger} - {2\pi}B(r)(1+{\sigma_z})
\end{equation}
It has the well known and interesting property to
 be exactly 
factorizable regardless of the shape of the magnetic field profile $B(r)$,
 i.e. 
it may be rewritten as the product of two first order differential operators 
(formally self-adjoint) $Q$ and $Q^\dagger$ such that $H= Q{Q^\dagger}$. 
This is a special example of a $N=2$ supersymmetric Hamiltonian\refnote{\cite{review}}.
 This feature is at 
the origin of the peculiar (non perturbative) topological properties of
 those systems.
For instance, Aharonov and Casher\refnote{\cite{ac}}
 did show explicitly that if the magnetic field is of 
finite flux, then the ground state degeneracy 
is $N-1$ where $N$ is the closest integer to the total flux 
(in units of $\Phi_0$). This is an example of the Atiyah Singer
 Index theorem\refnote{\cite{as}} which, for that case states that 
\begin{equation}
{\rm Index} Q = {\rm dim Ker} Q - {\rm dim Ker} {Q^\dagger} = N-1
\end{equation}
where ${\rm dim Ker} Q$ (resp. ${\rm dim Ker} {Q^\dagger}$) is the (finite) number of
 zero modes 
of $Q$ (resp. ${Q^\dagger}$) i.e. the number of solutions of the first order
 differential equation  $Q\Psi =0$ (resp. ${Q^\dagger}\Psi =0$).
 The index is an integer 
and as such a topological invariant in that sense that it remains unchanged
 under any classically
 permissible gauge transformation where for instance we may change randomly the
 profile
 of the magnetic field $B(r)$ but keeping unchanged the total magnetic flux.
 This results from the factorizability of the Pauli Hamiltonian a property
 which in
 general is not met by Schr\"{o}dinger Hamiltonians unless either
 we impose
 some special
 boundary conditions or we consider a uniform magnetic field $B$ where
 the corresponding 
Hamiltonian (Landau) is factorizable (see (3)) in terms of the operators
 $D$ and $
D^\dagger$ up to a constant $-{2\pi \over {\Phi_0}}B$, which sets the ground 
state energy (i.e. the lowest Landau level). In 
 this latter case, 
it is again possible (at least formally)
 to define an Index. It turns out to be infinite (and ill defined) 
which still corresponds to the infinite degeneracy of the lowest Landau level. 
Since the extensive (i.e. proportional to the surface) degeneracy of the
 ground state 
is an important ingredient for the Hall quantization in those systems, 
it might be tempting  
to preserve and extend the Index theorem to finite geometries. But then, 
it can be shown 
that any local choice of boundary conditions (e.g. Dirichlet or Neumann)
 destroys the factorizability property of the Schr\"{o}dinger Hamiltonian
 and therefore
 the condition of applicability of the Index theorem with this consequence
 that the corresponding 
ground state is always non degenerate.
 It was shown recently\refnote{\cite{aans}}
that the proper degeneracy as given by the Index can be
 restored using a special kind of non local boundary conditions.

To go further and relate these topological features to 
the magnetic response, we shall first focus on a specific example namely
 the case of 
a Aharonov-Bohm magnetic flux line.


\section{THE MAGNETIC RESPONSE OF AHARONOV-BOHM SYSTEMS}


We consider now the limiting case of a localized magnetic field of
 finite flux which 
corresponds to a Aharonov-Bohm flux line i.e. to a delta function magnetic
 field 
${\vec B}(r)= \Phi \delta ({\vec r}){\hat{e}_z}= {hc \over e}\phi 
\delta ({\vec r}){\hat{e}_z}$ and to a vector potential
 ${\vec A}({\vec r})= \phi{{ {\hat{e}_z} \times {\vec r}} \over {r^2}}$, where 
${\hat{e}_z}$ is the unit vector perpendicular to the plane. The corresponding 
Schr\"{o}dinger equation is obtained from Eq.(2) using
 $\phi(r)= \phi$. The angular 
momentum is a good quantum number and then the equation is separable.
 In each sector $m\in\integer$, a single valued solution of the radial equation is 
\begin{equation}
{\Psi_m}(kr)= a {J_{|m+\phi|}}(kr) + b {J_{-|m+\phi|}}(kr)
\end{equation}
where $E = {{{\hbar^2}{k^2}} \over 2m}$ is the energy, $a$ and $b$ are constants
 and 
${J_\nu}(kr)$ are Bessel functions. To describe an impenetrable flux line, we 
impose the boundary condition ${\Psi_m}(0)= 0$. Since ${J_{-|\nu|}}(kr)
{ \sim_{r \rightarrow 0}} (kr{)^{-|\nu|}}$
this amounts to take $b=0$ in order to have square integrable solutions
 at the origin.
 This choice 
is not as innocuous as it seems and we shall comment on it later.
 By choosing conveniently the normalization of the wavefunction we obtain
 $a=1$. 
Finally, in order to define completely our Hilbert space, we demand
 the two operators 
$D$ and $D^\dagger$ to be self adjoint such that for any states
 $f(r,\theta)$ and 
$g(r,\theta)$, $\langle f|Dg \rangle = \langle {D^\dagger}f|g \rangle$.
 The part 
$\phi \over r $ does not make any problem and by evaluating the radial
 integral on 
 a disk of radius $R$ (eventually $R \rightarrow \infty$), we obtain
\begin{equation}
R \int_{0}^{2\pi} d\theta {e^{i\theta}}{f^\ast}g{|_{r=R}} =0
\end{equation}
which is not fulfilled  in the general case (there is an exception using
 Dirichlet 
boundary conditions). A way to solve this problem is to translate the 
angular momentum of the eigenfunctions $f$ on the domain of the
 operator $D^\dagger$ by half a unit and therefore to consider the set
 ${J_{|m+{1 \over 2}+\phi|}}(kr) e^{i(m+ {1 \over 2})\theta}$.
Then, on the domain of the operator $D$, we consider functions $g$ of angular 
momentum decreased by half a unit. This amounts to consider a spinor like 
wavefunction, i.e. an effective Pauli Hamiltonian
 but where we keep only one of the two spin components.
 We shall come later to this using another point of view.

To characterize the magnetic response of the system,
 we calculate the (so called)
persistent current $I$ and the magnetization $\vec M$. Both are obtained
 from the
 local current density 
\begin{equation}
{\vec j}(r) = {\hbar \over m} {\rm Im} ({\Psi^\ast}({\vec \nabla} - 
{ie \over \hbar c}{\vec A})\Psi)
\end{equation}
where at zero temperature,
 we have to sum over all the occupied states up to the Fermi energy.
 Due to symmetry, only the azimuthal component $j_\theta$ is non zero such that
$I = \int_{0}^{\infty} dr {j_\theta}(r)$
and the total magnetic moment is 
${\vec M}= {\pi \over c} \int_{0}^{\infty} dr
 {r^2}{j_\theta}(r) {{\hat{e}}_z}$. 
From (8) we obtain 
\begin{equation}
{j_\theta}= \int_{0}^{k_f}{ kdk \over 2\pi} \sum_{m= -\infty}^{\infty}
{{m+ \phi +{1 \over 2}} \over r} {{J^2}_{|m+{1 \over 2}+\phi|}}(kr) 
\end{equation}
such that 
\begin{equation}
I= {{{k_f}^2} \over 8\pi} \sum_{m= -\infty}^{\infty} {\rm sgn}(m+{1 \over 2}+ \phi) 
\end{equation}
We first have to give a meaning to the divergent series $\eta \equiv \sum_{m}
{\rm sgn}(m+{1 \over 2}+ \phi)$. To that purpose, we first derive a
 relation between 
$\eta $ and the eigenvalues of the azimuthal current operator $\hat{J}= 
{i \over r} {\partial_\theta} + {{ \phi + {1 \over 2}} \over r }$.
 Consider the  
projection of $\hat{J}$ on a circle of radius $r=R$ (the exact value of $R$ is 
 irrelevant and taken to be one). The spectrum of the projected operator
 $\hat{J}_P$
 is $\lambda = m +{1 \over 2} + \phi$. Defining after Atiyah, Patodi and
 Singer\refnote{\cite{aps}} the quantity $\eta ({\hat{J}_P})= \sum_{\lambda < 0} 1 -
 \sum_{\lambda \geq 0} 1$, we can rewrite it 
\begin{eqnarray*}
\eta ({\hat{J}_P}) & = & {1 \over {\sqrt \pi}} \int_{0}^{\infty} 
{dt \over {\sqrt t}} Tr ({\hat{J}_P} e^{-t {{\hat{J}_P}^2}}) \\               
                   & = & \sum_{\lambda} {1 \over {\sqrt \pi}} {\rm sgn} 
(\lambda) 
 \int_{0}^{\infty} {dt \over {\sqrt t}} |\lambda| e^{-t |\lambda{|^2}} 
\end{eqnarray*}
which gives $\eta ({\hat{J}_P})= \eta$. It is of interest to rewrite
 $\eta ({\hat{J}_P})$ under the form 
\begin{equation}
\eta ({\hat{J}_P})= - {1 \over {2\sqrt \pi}} {\partial_\phi} \int_{0}^{\infty} 
{dt \over { t^{3 \over 2}}} Z(t,\phi)
\end{equation}
where $Z(t,\phi) = \sum_{m} e^{-t (m+{1 \over 2}+ \phi {)^2}} =
 Tr (e^{-t {{\hat{J}_P}^2}})$ is the partition function
 (or the Heat Kernel) at 
temperature $1 \over t$ of an electron constrained to move on 
a one-dimensional
 ring pierced 
 by a Aharonov-Bohm flux. Before calculating $\eta $ explicitly
 we notice that it was already considered in different contexts.
 To study the statistical properties of anyons, it was
 calculated\refnote{\cite{ouvry,comtet}} using a Feynman-Kac integral. In the context 
 of the Index theory of elliptic operators on manifolds with boundary, 
 it was calculated\refnote{\cite{aps2}} using a zeta function 
regularization i.e. by writing $\eta = \lim_{s \rightarrow 0}
 \sum_{\lambda \neq 0}
{\rm  sgn} (\lambda) |\lambda{|^{-s}}$.
 Here we shall evaluate it using the Poisson summation formula for the 
partition function $Z(t,\phi)$
\begin{eqnarray*}
Z(t,\phi) & = & \sum_{n=-\infty}^{\infty} \int_{-\infty}^{\infty} dx
e^{2i\pi nx} e^{-t (x+{1 \over 2} +\phi)^2} \\                 
          & = & \sqrt{\pi \over t} + 2{\sqrt \pi} \sum_{n  \neq 0}
 {1 \over \sqrt t}
e^{-{{\pi^2}{n^2}} \over t} e^{2i\pi n (\phi + {1 \over 2})}
\end{eqnarray*}
Inserting this expression in (11), we obtain
\begin{eqnarray*}
\eta ({\hat{J}_P}) & = & - {1 \over {2{\pi^2}}} {\partial_\phi}
 \sum_{n=1}^{\infty} 
{{{\sin^2}\pi n (\phi+{1 \over 2})} \over {n^2}} \\
                   & = & 2 \{\phi + {1 \over 2 }\} -1
\end{eqnarray*}
where $\{...\}$ represents the fractional part. 
The persistent current is then given by 
\begin{equation}
I = {{E_f} \over 2 } \eta ({\hat{J}_P})
\end{equation}
The amplitude of the total current 
is proportional to the Fermi energy ${E_f}$ and is therefore very large.
 This might be surprising since it is sometimes claimed that in the limit of 
an infinite system the normal persistent currents should vanish. This
 is indeed true for a one-dimensional ring of radius
 $R \rightarrow \infty$ but is incorrect in general.
 This point was discussed\refnote{\cite{aaas}} and we shall come back to it later
 when calculating 
the magnetic moment. We would like now to discuss the topological features
 of the current and relate it to the Index associated with the
 Aharonov-Bohm problem.
We saw previously that for a factorizable Hamiltonian the Index which counts 
the zero modes is defined by (5). Unfortunately, the Aharonov-Bohm
 Hamiltonian 
is at first sight non factorizable due to the additional factor $-2\pi B(r)$. 
However, by demanding self-adjointness of $D$ and $D^\dagger$, we ended up 
with a (factorizable) Pauli Hamiltonian. This result was obtained in
 a different (and perhaps more physical) way, noticing that 
 the energy spectrum 
 for the conveniently regularized problem\refnote{\cite{ouvry}}, 
is a non analytic function of the reduced flux $\phi$. This has 
its origin in the behaviour of the wavefunction for the angular
 momentum $m=0$. There, 
the unperturbed Hilbert space contains functions which do not vanish
 at the origin, 
while for $\phi \neq 0$, they do vanish like $r^{|\phi|}$. This gives rise to 
 singularities in perturbation theory which can be dealt with by
 adding a repulsive 
contact term in the Hamiltonian\refnote{\cite{ouvry}}.
To calculate the Index, we notice that since there is only one spin component 
(depending on the sign of the magnetic field), only one of the two operators $D$
 and $D^\dagger$ may have zero modes. 
The solutions of $Df =0$ are (for large $r$) of the form $f \propto r^{m-\phi - {1 \over 2}}$
 and their 
square integrability at infinity requires $m < {\phi} - {1 \over 2}$. To obtain well behaved 
solutions near the origin, we consider instead of the singular flux line a finite cylinder 
of radius $\epsilon$ in a uniform magnetic field $B$ such that the magnetic flux is 
$\Phi= B \pi {\epsilon^2}$. The zero modes inside this disk are solutions of 
$({\partial_r} + {i \over r} {\partial_\theta}+ {B \over 2}r)f=0$ which in the sector $m$ are 
of the form $ f(r) \propto {r^m}e^{-{{B{r^2}} \over 4}}$. It is straightforward to check that 
these solutions do match for $r= \epsilon$ where both inside and outside logarithmic derivatives 
are given by ${\partial_r}f{|_{r= \epsilon}}= {1 \over \epsilon}(m- \phi - {1 \over 2})f(\epsilon)$. 
Near $r=0$, the zero modes behave like $r^m$ so that square integrability requires $m \geq 0$. 
Finally, the number of zero modes i.e. the Index is given by ${\rm Index} = [\phi + {1 \over 2}]$, 
where $[...]$ represents the integer part. 
This corresponds to the degeneracy of the Aharonov-Bohm Hamiltonian (with its
 boundary condition).

 There is another way to obtain the Index which may shed some
 more light 
on its physical interpretation. To that purpose, we consider the scattering
 description
 of the  Aharonov-Bohm effect\refnote{\cite{ab}}. Berry et al.\refnote{\cite{berry}}
 proposed in that context to study the phase $\chi (r)$ 
 of the scattered part of the 
wavefunction and in particular the dislocations of the wavefronts
 defined as points 
where the modulus 
of the wavefunction vanishes\refnote{\cite{nye}}. The circulation ${1 \over {2 \pi}}
\oint_{c} d\chi= {1 \over {2 \pi}}\oint_{c} \vec{\nabla}\chi \cdot d\vec{r} $ of
 this 
phase over a close contour encircling the dislocation (i.e. the flux line)
 is an 
integer equals to $[\phi + {1 \over 2}]$
 i.e. to 
the Index. This result is not fortuitous and was clearly recognised in the
 mathematical literature\refnote{\cite{atiyah}} where the equivalence between the
 winding 
number around a singular point as introduced by Poincare\refnote{\cite{poincare}}, the
 degree of a 
 continuous map from the circle to the punctured complex plane and the 
Index of 
a conveniently defined elliptic operator was discussed. Each of those
 different points 
of view depends on the way we look at this problem. This way is geometric when 
considering instead of the initial map, say $f$, the mapping $f \over {|f|}$
 from the 
circle to the circle and count algebraically the number of intersections of
 the path 
with an arbitrary ray emanating from the origin. It is combinatorial when
 approximating 
the initial mapping by a piece-wise linear path and use combinatorial methods.
 It is
 differential in the way Berry et al.\refnote{\cite{berry}}
 considered it and analytic when studied 
from the point of view of the Index of an elliptic operator. It is the
 equivalence 
of those descriptions which is part of the richness of this problem.
 Although the
 differential point of view may appear at first sight more physical,
 the approach 
using the Index of an operator is more systematic which is useful in such non 
perturbative issues. 
Finally, comparing the different points of view, we arrive to the result
\begin{equation}
{\rm Index} = [\phi + {1 \over 2}]= {1 \over {2 \pi}}
\oint_{c}\vec{\nabla}\chi \cdot d\vec{r}
\end{equation}
Here, the first equality is obtained by calculating the
 zero modes 
of $D$, and the second 
comes from the definition of the azimuthal current 
density which tells us that the integral over a closed contour can be evaluated
either over 
a circle of large radius thus using the scattering form of the wavefunction\refnote{\cite{berry}} or 
on a circle of radius $r \rightarrow 0$ thus retaining in the series of  
Bessel functions (9) only the lowest indices which corresponds
 to a calculation of the zero modes.
 The Index as a function of the flux $\phi$ shows plateaus and jumps at 
half integers.
  As explained by Berry et al.\refnote{\cite{berry}}, these jumps 
correspond to a long range reorganisation of the 
wavefronts in contrast to the behaviour at integers which describes local 
changes of the wavefronts around the flux line.
It might be interesting at this stage to compare the previous results with 
those obtained in the context of superfluids or superconductors\refnote{\cite{pgdg}}.
 For superfluids, the gradient of the phase of the macroscopic
 Onsager-Feynman wavefunction measures the superfluid velocity and 
then the total current. Being a gradient, the latter describes an irrotational 
flow such that the circulation of the velocity on a closed curve is quantized 
as observed experimentally by Vinen\refnote{\cite{vinen}}. As an outcome, the 
force exerted on a body immersed in the fluid vanishes (d'Alembert paradox). 
For a superconductor, the bulk current density vanishes (Meissner effect),
 such that from the circulation on a closed curve we obtain the quantization
 of the magnetic flux. But this is different from the Index theorem 
which states that the circulation on a closed curve of 
 the gradient of the phase which must be an integer, depends on the flux 
as given by (13).

Considering the sum of $\eta ({\hat{J}_P})$
 and of the Index, we obtain the relation
\begin{equation}
{\rm Index} + \eta ({\hat{J}_P})= \phi
\end{equation}
which in other words gives
 a sum rule between the radial and azimuthal integrals of the current 
density $j_{\theta}$. This relation is the 
expression of a general result\refnote{\cite{aps}} which generalises to non compact 
spaces the Atiyah Singer Index theorem.

It might be also of interest to rewrite (14) using for ${I \over {E_f}}$ 
its expression in terms of scattering phase shifts\refnote{\cite{aaas}},
 ${I \over {E_f}}={1 \over \pi}{\partial_\phi} \sum_{m}\delta_{m}(\phi)$
 where the phase shifts for a 
spin one half are given \refnote{\cite{hagen}} by  
$\delta_{m}(\phi)= {\pi \over 2}(|m+\phi + {1 \over 2}|- |m|)$. Then,
\begin{equation}
{\rm Index} +{1 \over \pi}{\partial_\phi} \sum_{m}\delta_{m}(\phi) = \phi
\end{equation}
Under this form, it is easy to see that the phase shift term represents the 
boundary contribution to the Index theorem for non compact manifolds.

 The relation between the persistent currents and the scattering phase shifts 
suggests an interesting analogy between this problem and the screening of 
an electric charge as given by the Friedel sum rule\refnote{\cite{friedel}}. In the 
latter case, inserting an external charge $Z$ in a metal the electrical 
 neutrality is expressed self-consistently by the sum rule 
$\pi Z =\delta({E_f})$ where $\delta({E_f})$ is the total scattering phase 
shift calculated at the Fermi energy 
 describing the scattering of electrons by the external 
charge. For the Aharonov-Bohm case, we can interpret the persistent currents
 in a 
similar way saying that the magnetic flux line is screened by current 
loops of electrons sitting at infinity. Using (15), the current can be 
expressed in terms of a topological invariant, namely the Index. 
The Friedel sum rule may be understood similarly.
 The total phase shift which 
in principle depends on the microscopic details of the potential created by 
the external charge is in fact
 a function of $Z$ only, irrespective of the way this charge is distributed.

Finally, we need to evaluate the total magnetic moment. 
Substituting (9) into the Biot-Savart law gives 
\begin{equation}
{M}= \mu_{B} \int_{0}^{R} {r^2 dr} \int_{0}^{k_f}{ kdk } 
\sum_{m= -\infty}^{\infty}
{{m+ \phi +{1 \over 2}} \over r} {{J^2}_{|m+{1 \over 2}+\phi|}}(kr).
\end{equation}
 Using the Euler-Maclaurin summation formula\refnote{\cite{prange}} we 
obtain an asymptotic 
expansion in terms of the large 
parameter $k_f R$ such that the magnetic moment can be written as a series 
$ \sum_{m= -\infty}^{\infty}
F({m+ \phi +{1 \over 2}}) $, where the function $F$ is 
\[
F(x)= x \mu_{B} \int_{0}^{R} {r dr} \int_{0}^{k_f}{ kdk } 
 {{J^2}_{|x|}}(kr).
\]
A naive application of the Euler-Maclaurin summation formula would give zero 
 due to the vanishing of the function $F(x)$ at plus and minus  
infinity, for all finite $k_f$ and $R$. However this 
function is singular at $x=0$, where its second derivative is discontinuous
with a jump equal to $-2\mu_{B}$ (this result has small corrections of the 
order of $\mu_{B}/k_f R$). Then, the usual derivation of the 
Euler-Maclaurin summation formula\refnote{\cite{bromwich}} should be 
revised, taking into account the singularity at $m+ \phi +{1 \over 2}=0$.
Finding an integer $m_1$ such that $m_1+ \phi +{1 \over 2} < 0 < 
m_1+ \phi +{3 \over 2}$, it turns out that the important parameter (which 
determines the position of the singularity) in the derivation of the 
modified Euler-Maclaurin summation formula is $m_1+ \phi +{3 \over 2}$,
which is equal to the fractional part of the total flux $\phi$. One can 
prove then that the correction itself is proportional to the jump of the second
derivative times the Bernoulli polynomial of third order, evaluated at 
$\{ \phi \}$. More precisely 
 we obtain
\begin{equation}
M \sim  {\mu_{B} \over 3} B_3(\{\phi \}) + {\rm O}(\mu_{B}/k_f R),
\end{equation}
where $B_3(x)=x^3-3 x^2/2+ x/2$ is a Bernoulli polynomial. The magnetization
of the system, which is according to the definition the magnetic moment per 
unit area, vanishes in the thermodynamic limit. Indeed, (17) corresponds 
to the finite magnetic moment of an infinite system. 
Therefore, even though the current is large, the magnetic 
response is experimentally inaccessible.

\section{HEAT KERNEL AND PARTITION FUNCTION FOR THE AHARONOV-BOHM PROBLEM}


From (11), we obtained an expression for the persistent current $I$
 in terms of an integral over the partition function $Z(t,\phi)$ of
 an electron moving on a one-dimensional ring pierced by a Aharonov-Bohm flux
 at an effective temperature $1 \over t$. On the other hand, since the 
persistent current is a thermodynamic quantity, it is possible in principle 
to express it in terms of the Heat Kernel (or the partition function)
$P(\beta, \phi)= Tr (e^{-\beta H(\phi)})$ where $ H(\phi)$ is the  
Hamiltonian defined by (2) for a constant $\phi$ and where
 $\beta$ is the inverse
 temperature. 
To obtain a relation between these two quantities, we start writing the 
thermodynamic grand potential $\Omega(\beta,\phi)= \int_{0}^{\infty}dE 
f(E,\mu,\beta)N(E)$ where $f$ is the Fermi-Dirac distribution function,
 $\mu$ the 
chemical potential and $N(E)$ the integrated density of states (the counting
 function). The current is then given by $I= - {{\partial\Omega} \over 
{\partial\phi}}$ in the appropriate units. We then define after 
Sondheimer and Wilson\refnote{\cite{sw}} the 
function $z(E,\phi)$ by
\begin{equation}
{P(\beta, \phi) \over {\beta^2}}=  \int_{0}^{\infty}dE e^{-\beta E}z(E,\phi)
\end{equation}
The grand potential as a function of $z(E,\phi)$ rewrites $\Omega(\beta,\phi)=
 \int_{0}^{\infty}dE {{\partial f} \over {\partial E}}z(E,\phi)$. In the 
asymptotic limit $\beta \mu \gg 1$, i.e. for very low temperatures we obtain 
$I={ {\partial z(\mu,\phi)} \over {\partial \phi}}$ from which we deduce   
\begin{equation}
{1 \over {\beta^2}}{{\partial P(\beta, \phi)} \over {\partial \phi}}
=  \int_{0}^{\infty}dE e^{-\beta E}I(E,\phi)
\end{equation}
By inserting (11) for $I$, we obtain
\begin{equation}
{{\partial P(\beta, \phi)} \over {\partial \phi}}
=  - {1 \over {4\sqrt \pi}} {\partial_\phi} \int_{0}^{\infty} 
{dt \over { t^{3 \over 2}}} Z(t,\phi)
\end{equation}
which relates the partition function $P(\beta, \phi)$
 of the two-dimensional problem to the partition 
function $Z(t,\phi)$ of an effective one-dimensional Aharonov-Bohm problem.


\section{BEYOND THE AHARONOV-BOHM CASE}


We would like now to discuss the extension of the previous results to more 
general magnetic field configurations. We noticed that
 the Pauli Hamiltonian (4) is always factorizable irrespective of the field 
configuration $B(r)$. Using this property, the ground state degeneracy is 
given by the Index. In addition to that 
result we obtained for the case of a flux line a relation
(14) between the Index and $\eta ({\hat{J}_P})$. This relation holds for 
any factorizable Hamiltonian. Is it possible to interpret it in terms of the 
magnetic response of the system?  To that purpose we
 consider once again the simpler case of a uniform magnetic 
field. There, the orbital and spin parts of the magnetization can be separated. 
In the radial gauge $\vec{A}= {B \over 2}(-y,x)$, the Pauli Hamiltonian is 
$H= {1 \over {2m}}(\vec{p} -{e \over c}\vec{A}{)^2} - {\mu_B}B{\sigma_z} 
\equiv {H_0}  - {\mu_B}B{\sigma_z}$. This can be written in a matricial form 
\begin{eqnarray*}
 H =  \left( \begin{array}{ccc} 
{H_+}  & & {\mbox{$0$}} \\ 
{0} & & \mbox{ ${H_-}$} 
\end{array}
 \right) 
\end{eqnarray*}
where $H_{\pm}= {H_0} {\pm}{\mu_B}B$. Defining the supercharge 
$Q = {1 \over {\sqrt {4m}}} \vec{\Pi}.\vec{\sigma}=  {1 \over {\sqrt {4m}}}
({\Pi_x}{\sigma_x}+ {\Pi_y}{\sigma_y} )$ where 
${\Pi_x}= {p_x}- {e \over c}{A_x}$ and ${\Pi_y}= {p_y}- {e \over c}{A_y}$,
we can write the Hamiltonian under the supersymmetric form $H= 2{Q^2}$. 
The associated creation and annihilation
 operators $A$ and $A^\dagger$ are given by $A= {1 \over {\sqrt {2m}}} 
({\Pi_x}-i{\Pi_y})$ and ${A^\dagger}= {1 \over {\sqrt {2m}}} 
({\Pi_x}+i{\Pi_y})$, such that ${H_+}= A{A^\dagger}$ and ${H_-}= {A^\dagger}A$.
 According to (5), the Index is then given\refnote{\cite{simon}} by ${\rm Index}= {\rm dim Ker} A - 
{\rm dim Ker} {A^\dagger}$. To go further and connect it with 
the thermodynamics of the system, we notice that ${\rm Ker} A = {\rm Ker} {A^\dagger}A 
= {\rm Ker} {H_-}$ and ${\rm Ker}{A^\dagger}= {\rm Ker}  A{A^\dagger}= 
{\rm Ker}{H_+}$.
 Since the spectra of ${H_+}$ and ${H_-}$  are identical except perhaps 
for the zero modes, it is possible to rewrite 
the Index using the regularization
\begin{equation}
{\rm  Index} = \lim_{t \rightarrow 0} {\rm Tr} ( e^{-t{H_+}}-  e^{-t{H_-}}) = 
 \lim_{t \rightarrow 0}{\rm Tr} ({\sigma_z} e^{-tH})
\end{equation}
The spectrum of the Pauli 
Hamiltonian is given as for the Landau case by a set of infinitely degenerate 
Landau levels. The supersymmetric pairing of all the excited states means that 
they do not participate to the spin magnetization since there are no states 
with respectively spin up and spin down which are unpaired on an excited 
Landau level. Then, a finite spin magnetization stems only from possible 
unpaired states in the lowest Landau level. For a non 
interacting electron gas, the Fermi energy $E_f$ is determined by fixing 
the total number $N$ of electrons through $N= {\rm Tr} (\Theta ( 
{E_f}- H))$. The spin magnetization can be written ${M_S}= {\mu_B}
({N_+}-{N_-})$ where ${N_+}$ and ${N_-}$ are the number of electrons with spin 
up and spin down respectively. Using a regularization
 equivalent to (21) we obtain the result 
\begin{equation}
{M_S}= {\mu_B} {\rm Tr}({\sigma_z}
\Theta({E_f} - {H}))=  {\mu_B}{\rm Index}
\end{equation}
In order to calculate explicitly the Index, we write the traces in (21)
\begin{equation}
{\rm Index} = \lim_{t \rightarrow 0} \int {d^2}x \{ \langle x| e^{-t{H_+}}|x
\rangle -  \langle x| e^{-t{H_-}}|x\rangle \}
\end{equation}
The Heat Kernels for respectively spin up and spin down 
which do appear in (23) are given for the 
case of a uniform magnetic field by 
\begin{equation}
\int {d^2}x  \langle x| e^{-t{H_{\pm}}}|x
\rangle = {1 \over 2}{{\Phi} \over {\Phi_0}}e^{{\pm}{{\hbar {\omega_c}t}
 \over 2}}{ 1 \over {\sinh ({{\hbar {\omega_c}t}
 \over 2}})}
\end{equation}
 where ${\omega_c}= {eB \over mc}$ is the cyclotron frequency.
 The Index is effectively $t$-independent and given by 
the total magnetic flux (in units of the flux quantum $\Phi_0$) and is
 therefore infinite. 

The outcome of this is the relation between the Index and the spin
 magnetization given by (22). As anticipated from the role of the 
Zeeman term for the Pauli Hamiltonian to be factorizable, the Index 
appears as a measure of the spin magnetization. Nevertheless, this result is 
very peculiar and applies only for the case of a uniform magnetic field, 
where the orbital and Zeeman parts in the Hamiltonian do commute, such 
that the respective  magnetic responses are independent. In the general case 
of a inhomogeneous magnetic field, it is not anymore possible to disentangle 
those two parts. Another peculiarity of the uniform field case is 
that the Index is infinite and formally ill-defined. For a  magnetic
 field of finite flux, it is not anymore the case and
 the total magnetic response is still given by (14) as a 
consequence of the Atiyah, Patodi, Singer\refnote{\cite{aps}} theorem.
  

\section{CONCLUSION}


We have presented a study of the magnetic response of a degenerate 
non interacting electron gas in a inhomogeneous magnetic field from the 
point of view of the Index Theory. For the limiting case of a 
Aharonov-Bohm flux line, it allowed us to find a relation 
between the persistent current and a topological invariant, namely the Index 
of an operator and to calculate it. We obtained a relation between this 
Index and the winding number defined geometrically from the behaviour of 
the wavefunction. To provide another physical interpretation of the Index, 
we showed that for the case of a uniform magnetic field, it is proportional 
to the spin magnetization and we presented a general regularization scheme 
to calculate it from the partition function (or the Heat Kernel). This 
connection to the magnetic response can be generalised to other systems as 
an application of the Atiyah, Patodi, Singer Theorem. The extension of this 
point of view to study transport coefficients in the Quantum Hall 
regime is appealing. It allows\refnote{\cite{an}} to relate the Hall conductance 
to a suitably defined Index for systems with a boundary. This 
generalises other approaches using topological numbers\refnote{\cite{as2,thouless}}. 
Among other problems where the point of view presented here may 
be relevant, is the puzzle of mesoscopic persistent currents in small 
(but multichannel) rings.
 There, as discussed by Leggett\refnote{\cite{leggett}}, the existence of nodal lines 
where the wavefunction vanishes would involve more sophisticated topological 
tools in order to make some progress.

\subsection{Acknowledgements} 
This work is supported in part by a grant from the Israel 
Academy of Sciences and by the fund for promotion of Research at the Technion.
It is a pleasure to thank M. Atiyah, J. E. Avron, M. Berry and K. Mallick
 for very useful 
discussions. R. N. also acknowledges the support of the NSF grant DMR 9625549 
and thanks R. E. Prange for enlightening discussions.

\end{document}